\documentclass{article}
\usepackage{times}
\usepackage{amsfonts}
\usepackage{graphicx}

\begin{document}

\noindent
{\Large ON PADMANABHAN'S DUALITY INVARIANCE\\ AND THE QUANTUM OF LENGTH}\\
\vskip1cm
\noindent
{\bf P. Fern\'andez de C\'ordoba}$^{1}$, {\bf J.M. Isidro}$^{2}$ and {\bf A. Pereira Garc\'{\i}a}$^{3}$\\
Instituto Universitario de Matem\'atica Pura y Aplicada,\\ Universitat Polit\`ecnica de Val\`encia, Valencia 46022, Spain.\\
$^{1}${\tt pfernandez@mat.upv.es}, $^{2}${\tt joissan@mat.upv.es},\\ $^{3}${\tt apergar2@posgrado.upv.es}
\vskip.5cm
\noindent
\vskip.5cm
\noindent
{\bf Abstract} We provide a field--theoretic construction of Padmanabhan's duality--invariant Feynman propagator for a massive point particle in Euclidean space $\mathbb{R}^D$. Padmanabhan's propagator includes quantum--gravity effects due to the existence of a quantum of length $\ell$. Including $O(\ell^2)$ corrections, the corresponding field--theory model turns out to be a free, massive scalar defined in $\mathbb{R}^{D+2}$. The two additional dimensions with respect to the original $\mathbb{R}^D$ provide the necessary room, so to speak, for quantum--gravity fluctuations.

\tableofcontents

\section{Introduction}\label{sekt1}

Let us consider $D$--dimensional Euclidean space $\mathbb{R}^D$, on which we will study the propagation of a relativistic point particle and also of a scalar quantum field.

\subsection{A relativistic point particle}\label{sekt2}

A free, massive scalar particle is described classically by an action integral 
\begin{equation}
m{\cal S}_D\left[p({\bf x}, {\bf y})\right]=m\int_{\bf x}^{\bf y}{\rm d}s,
\label{linint}
\end{equation}
the integral being computed along the path $p({\bf x}, {\bf y})$ connecting ${\bf x}\in\mathbb{R}^D$ to ${\bf y}\in\mathbb{R}^D$. Quantum--mechanically, the sum over all paths from ${\bf x}$ to ${\bf y}$, 
\begin{equation}
{\cal G}_D({\bf x}, {\bf y})=\sum_{{\rm paths}\,p}\exp\left\{-m{\cal S}_D\left[p({\bf x}, {\bf y})\right]\right\},
\label{treintayuno}
\end{equation}
defines the Euclidean Feynman propagator. It turns out to depend on the points ${\bf x}, {\bf y}$ through the geodesic distance between them. 

In quantum gravity one is interested in constructing a Feynman propagator that is invariant under the duality transformation \cite{PADDY0, PADDY1}
\begin{equation}
{\cal S}_D\longrightarrow\frac{L^2}{{\cal S}_D}.
\label{sesenta}
\end{equation}
Above, $L$ is a {\it quantum of length}\/ on spacetime. To this end one modifies the path integral (\ref{treintayuno}) to the manifestly duality--invariant path integral
\begin{equation}
{\cal G}_D^{{\rm (QG)}}({\bf x}, {\bf y})=\sum_{{\rm paths}\,p}\exp\left(-m\left\{{\cal S}_D\left[p({\bf x}, {\bf y})\right]+\frac{L^2}{{\cal S}_D\left[p({\bf x}, {\bf y})\right]}\right\}\right).
\label{treintaydos}
\end{equation}
It has been established in refs. \cite{PADDY0, PADDY1} that the propagator (\ref{treintaydos}) includes lowest--order, quantum--gravitational effects due to the presence of  the zero--point  length $L$. The superscript ``QG" stands for {\it quantum gravity}\/, and refers to the presence of $L\neq 0$.\footnote{Whenever Feynman propagators depend on ${\bf x}, {\bf y}\in\mathbb{R}^D$ through the Euclidean norm $s=\sqrt{({\bf x}-{\bf y})^2}$, we will denote them interchangeably as ${\cal G}_D(s)$, ${\cal G}_D^{\rm (QG)}(s)$ or as ${\cal G}_D({\bf x}, {\bf y})$, ${\cal G}_D^{\rm (QG)}({\bf x}, {\bf y})$; we will often set ${\bf y}={\bf 0}$ and just write ${\cal G}_D({\bf x})$, ${\cal G}_D^{\rm (QG)}({\bf x})$, and analogously for momentum--space propagators. The same conventions will apply to the propagators $G_D$ and $G_D^{\rm (QG)}$ to be defined presently.}

The path integral (\ref{treintayuno}), although nonquadratic, has been computed exactly with the result \cite{PADDY0, PADDY1}
\begin{equation}
{\cal G}_D(s)=\frac{m^{D-2}}{(2\pi)^{D/2}}\frac{K_{D/2-1}\left(ms\right)}{\left(ms\right)^{D/2-1}}, \qquad s^2=({\bf x}-{\bf y})^2. 
\label{catorce}
\end{equation}
Above, $K_n(z)$ is a MacDonald function.\footnote{In our use of special functions we follow the conventions of ref. \cite{LEBEDEV}.} 

In the presence of a quantum of length $L\neq 0$, the duality--invariant propagator ${\cal G}_D^{\rm (QG)}$ in (\ref{treintaydos}) has been found in refs. \cite{PADDY0, PADDY1} to be
\begin{equation}
{\cal G}_D^{\rm (QG)}(s)=\frac{m^{D-2}}{(2\pi)^{D/2}}\frac{K_{D/2-1}\left(m\sqrt{s^2+4L^2}\right)}{\left(m\sqrt{s^2+4L^2}\right)^{D/2-1}}.
\label{kagonesteve}
\end{equation}
We observe that the functional form of the propagator (\ref{kagonesteve}) is the same as in Eq. (\ref{catorce}), the only difference being the shift 
\begin{equation}
s=\sqrt{s^2}\longrightarrow\sqrt{s^2+4L^2}
\label{chif}
\end{equation}
due to the zero--point length $L>0$.  For notational convenience we will introduce 
\begin{equation}
\ell=2L,
\label{40}
\end{equation}
which will make numerous equations look neater, {\it e.g.}\/ the shift (\ref{chif})
\begin{equation}
s=\sqrt{s^2}\longrightarrow\sqrt{s^2+\ell^2}.
\label{chof}
\end{equation}
We have initially used the notation $L$ for the quantum of length, as that is the convention followed in refs. \cite{PADDY0, PADDY1}, but we will henceforth use $\ell$ as defined in (\ref{40}). In particular we can rewrite the propagator (\ref{kagonesteve}) as
\begin{equation}
{\cal G}_D^{\rm (QG)}(s)={\cal G}_D\left(\sqrt{s^2+\ell^2}\right).
\label{98}
\end{equation}

\subsection{A quantum field}\label{sekt3}

The quantum theory of a relativistic scalar field in Euclidean $\mathbb{R}^D$ can be described by the action\footnote{We stress the difference between the action integrals (\ref{linint}) and (\ref{508}) by using the different fonts ${\cal S}$ and $S$. The corresponding Feynman propagators are denoted ${\cal G}$ and $G$, respectively. The point--particle action ${\cal S}$ has mass dimension $-1$; the field--theory action $S$ is dimensionless.}
\begin{equation}
S_D\left[\Phi^*,\Phi\right]=\int_{\mathbb{R}^D}{\rm d}^D{\bf x}\,\Phi^*({\bf x})\left(-\nabla^2+m^2\right)\Phi({\bf x}).
\label{508}
\end{equation}
The corresponding Feynman propagator $G_D(s)$ can be obtained starting from the Fourier decomposition\footnote{Our conventions for Fourier transforms on $\mathbb{R}^D$:
$$
{\cal F}[f]({\bf p})=\hat f({\bf p})=\int_{\mathbb{R}^D}{\rm d}^D{\bf x}\,f({\bf x}){\rm e}^{-{\rm i}{\bf p}\cdot{\bf x}},\qquad
{\cal F}^{-1}[\hat f]({\bf x})=f({\bf x})=\int_{\mathbb{R}^D}\frac{{\rm d}^D{\bf p}}{(2\pi)^D}\,\hat f({\bf p}){\rm e}^{{\rm i}{\bf p}\cdot{\bf x}}
$$
} 
\begin{equation}
G_D(s)=\int_{\mathbb{R}^D}\frac{{\rm d}^D{\bf p}}{(2\pi)^D}\frac{\exp\left[{\rm i}{\bf p}\cdot\left({\bf x}-{\bf y}\right)\right]}{p^2+m^2},  \qquad s^2=({\bf x}-{\bf y})^2, \qquad p^2={\bf p}^2,
\label{trescientoscuatro}
\end{equation}
and inserting the Schwinger proper--time representation of $(p^2+m^2)^{-1}$:
\begin{equation}
\frac{1}{p^2+m^2}=\int_0^{\infty}{\rm d}\tau\,\exp\left[-\tau(p^2+m^2)\right].
\label{trescientoscinco}
\end{equation}
Substituting (\ref{trescientoscinco}) into (\ref{trescientoscuatro}), exchanging the order of integration and computing the Gaussian in the first place yields
$$
G_D(s)=\frac{1}{(4\pi)^{D/2}}\int_0^{\infty}{\rm d}\tau\,\tau^{-D/2}\exp\left(-m^2\tau-\frac{s^2}{4\tau}\right)
$$
\begin{equation}
=\frac{m^{D-2}}{(2\pi)^{D/2}}\frac{K_{D/2-1}\left(ms\right)}{\left(ms\right)^{D/2-1}}={\cal G}_D(s).
\label{cientoveintuno}
\end{equation}
So one has the equality ${\cal G}_D(s)=G_D(s)$ between the Feynman propagators computed with respect to the point--particle action ${\cal S}_D$ and the field--theory action $S_D$.

It is the purpose of this article to construct a field--theoretic counterpart to the point--particle theory described by the self--dual action integral in Eq. (\ref{treintaydos}).

\section{An $O(\ell^2)$ action for quantum--gravity corrections}

In ref. \cite{ISIDRO} we have established that, for a point particle described by the action functional in Eq. (\ref{treintaydos}), the following identity holds:
\begin{equation}
{\cal G}^{\rm (QG)}_{D}(s)={\cal G}_{D}\left(\sqrt{s^2+1}\right)=\sum_{n=0}^{\infty}\frac{(-\pi)^n}{n!}{\cal G}_{D+2n}(s).
\label{99}
\end{equation}
The propagators ${\cal G}_D(s)$ and ${\cal G}^{\rm (QG)}_{D}(s)$  are as in Eqs. (\ref{catorce}) and (\ref{kagonesteve}), respectively, upon setting $m=1$, $2L=\ell=1$ in the latter. On the left--hand side above, the variable $s=\sum_{j=1}^Dx_j^2$ stands for the Euclidean distance in $\mathbb{R}^D$. On the right--hand side, $s=\sum_{j=1}^{D+2n}x_j^2$ denotes the Euclidean distance in $\mathbb{R}^{D+2n}$. However, $s$ in the right--hand side is such that its numerical value coincides with that of $s$ on the left--hand side. As explained in ref. \cite{ISIDRO}, Eq. (\ref{99}) is the Taylor expansion of ${\cal G}_{D}\left(\sqrt{s^2+\ell^2}\right)$ in powers of $\ell$, after setting $\ell=1$. By parity, only even powers of $\ell$ appear in the expansion; this accounts for the summation being over all dimensions $D+2n$.

The interpretation of Eq. (\ref{99}) reads: QG effects on point--particle propagators in $\mathbb{R}^D$ arise from an infinite summation over all {\it QG--free}\/ propagators in all higher dimensions $D+2n$. Since the ${\cal G}_{D+2n}(s)$ on the right--hand side are free of QG effects, one can equate point--particle propagators to field--theory propagators and replace
${\cal G}_{D+2n}(s)$ with $G_{D+2n}(s)$. Thus (\ref{99}) becomes
\begin{equation}
{\cal G}^{\rm (QG)}_{D}(s)={\cal G}_{D}\left(\sqrt{s^2+1}\right)=\sum_{n=0}^{\infty}\frac{(-\pi)^n}{n!}G_{D+2n}(s),
\label{100}
\end{equation}
while the two ${\cal G}_D$'s on the left--hand side bear the imprint of QG effects through the quantum of length $\ell=1$. But we are not allowed to replace 
${\cal G}^{\rm (QG)}_{D}(s)$ with $G^{\rm (QG)}_{D}(s)$, because we do not know $G^{\rm (QG)}_{D}(s)$ yet. In fact we still lack an action functional 
$S^{\rm (QG)}_{D}$ from which to derive the propagator $G^{\rm (QG)}_{D}(s)$.

A clue as to how to determine $S^{\rm (QG)}_{D}$ comes from the following observation. Truncation of the expansion (\ref{100}) to the first two terms $n=0$, $n=1$
\begin{equation}
{\cal G}_D^{\rm (QG)}(s)=G_D(s)-\pi\ell^2G_{D+2}(s)+O(\ell^4).
\label{101}
\end{equation}
suggests that the sought--for, field--theory action functional $S_D^{\rm (QG)}$ should be given by 
\begin{equation}
S_D^{\rm (QG)}=S_D-\pi m^2\ell^2S_{D+2}+O(\ell^4).
\label{93}
\end{equation}
We have made an educated guess for the coefficient $-\pi m^2\ell^2$: the factor $-\pi$ is determined by the $n=1$ term in (\ref{99}), the factor $m^2$ is required to cancel dimensions against $\ell^2$ in (\ref{101}), and we set $m=1$, $\ell=1$ at the end. 

The linear combination (\ref{93}) requires some care.  $S_{D+2}$ is a functional of the field $\Phi_{D+2}:\mathbb{R}^{D+2}\rightarrow\mathbb{C}$. The latter must depend on the variables $x_1,\ldots, x_{D+2}$ in such a way that the last two of them, $x_{D+1}$, $x_{D+2}$, always satisfy \cite{PADDY6}
\begin{equation}
x_{D+1}^2+x_{D+2}^2=\ell^2.
\label{104}
\end{equation}
The additional 2 dimensions $x_{D+1}$, $x_{D+2}$ in $\mathbb{R}^{D+2}$ with respect to $\mathbb{R}^D$ play the role of an extension, along which the quantum fluctuations encoded in the $D$--dimensional quantum of length $\ell$ can play out. The propagator $G_D^{\rm (QG)}(s)$ corresponding to the ansatz (\ref{93}) is
\begin{equation}
G_D^{\rm (QG)}(s)=G_D(s)-\pi\ell^2 G_{D+2}(s)+O(\ell^4).
\label{102}
\end{equation}
{}Finally comparing Eqs. (\ref{102}) and (\ref{101}) leads to the equality between field--theory and point--particle propagators
\begin{equation}
G_D^{\rm (QG)}(s)={\cal G}_D^{\rm (QG)}(s) + O(\ell^4),
\label{103}
\end{equation}
valid up to, and including, $O(\ell^2)$ corrections in QG.

A word of clarification is in order regarding the physical status of the field $\Phi_{D+2}$. The linear combination of actions (\ref{93}) should not be interpreted as a fundamental theory describing physical degrees of freedom propagating in a higher--dimensional spacetime. Instead, it is merely a local, effective mathematical representation designed to capture the nonlocal features induced by the quantum of length $\ell$ in the physical space $\mathbb{R}^D$. In this sense, the two auxiliary dimensions $x_{D+1}$, $x_{D+2}$ play a role analogous to the Schwinger proper--time parameter $\tau$ in Eq. (\ref{trescientoscinco}). They provide a geometric setting that allows us to bypass the infinite tower of spatial derivatives typically found in UV--modified theories, preserving second--order field equations at the expense of an effective dimensional extension. The constraint (\ref{104}) ensures that the Feynman propagators correctly inherit the duality--invariant quantum gravity corrections up to ${\cal O}(\ell^2)$ on the physical manifold $\mathbb{R}^D$.

\section{Why a naive discretisation of $\mathbb{R}^D$ fails}\label{sekt6}

The simple example of a naive discretisation of $\mathbb{R}^D$ presented below fails because it does not reproduce the right $O(\ell^2)$ corrections due to a quantum of length. Admittedly, the failure of one given discretisation does not imply all other discretisations will fail. Nevertheless, the following is an instructive exercise to work out.

\subsection{Field--theory action functional on a lattice}\label{sekt4}

Let us define the functional
\begin{equation}
S_{\ell}\left[\Phi^*,\Phi\right]=\sum_{{\bf x}\in\mathbb{Z}^D(\ell)}\Phi^*({\bf x})\left(-\nabla_{\ell}^2+\frac{m^2}{2}\right)\Phi({\bf x}),
\label{506}
\end{equation}
where
\begin{equation}
\mathbb{Z}^D(\ell)=\left\{\ell(n_1,\ldots, n_D),\quad n_j\in\mathbb{Z},\quad j=1,\ldots D\right\}
\label{506b}
\end{equation}
is a $D$--dimensional integral lattice. If the ${\bf e}_j$, $j=1,\ldots, D$, denote the canonical basis vectors of the space $\mathbb{R}^D$, then the ${\bf e}_j$ span the lattice $\mathbb{Z}^D$, while the $\ell{\bf e}_j$ span the lattice $\mathbb{Z}^D(\ell)$ summed over in (\ref{506}). There is a natural definition of a Laplacian operator $\nabla_{\ell}^2$ on $\mathbb{Z}^D(\ell)$:
\begin{equation}
\nabla_{\ell}^2\Phi({\bf x})=\frac{1}{\ell^2}\sum_{j=1}^D\left[\Phi({\bf x}+\ell{\bf e}_j)-2\Phi({\bf x})+\Phi({\bf x}-\ell{\bf e}_j)\right].
\label{506c}
\end{equation}
As $\ell\to 0$ the lattice $\mathbb{Z}^D(\ell)$ gives back the continuum space $\mathbb{R}^D$, and (\ref{506c}) reduces to the standard Laplacian thereon.

Now let $\ell_0$ denote any fixed length scale, {\it e.g.}\/, $\ell_0=m^{-1}$, and define
\begin{equation}
\tilde\ell=\frac{\ell_0^2}{\ell}.
\label{80}
\end{equation}
It would seem natural to conjecture that a point particle in the presence of a quantum of length $\ell$, as described by the self--dual action in Eq. (\ref{treintaydos}), should possess a field--theory counterpart given by
\begin{equation}
S^{\rm (QG)}\left[\Phi^*,\Phi\right]=S_{\ell}\left[\Phi^*,\Phi\right]+S_{\tilde\ell}\left[\Phi^*,\Phi\right]
\label{81}
\end{equation}
$$
=\sum_{{\bf x}\in\mathbb{Z}^D(\ell)}\Phi^*({\bf x})\left(-\nabla_{\ell}^2+\frac{m^2}{2}\right)\Phi({\bf x})+\sum_{\tilde {\bf x}\in\mathbb{Z}^D(\tilde\ell)}\Phi^*(\tilde{\bf x})\left(-\nabla_{\tilde{\ell}}^2+\frac{m^2}{2}\right)\Phi(\tilde{\bf x}).
$$
{\it Yet the above conjecture is wrong}\/. Before proving it wrong, some remarks are in order. First, the model defined by Eq. (\ref{81}) is not a continuum, but a lattice, field theory. Second, two different Laplacian operators are involved: $\nabla_{\ell}^2$ and $\nabla^2_{\tilde{\ell}}$. Third, the action functional (\ref{81}) is manifestly invariant under the transformation
\begin{equation}
\ell\rightarrow\tilde\ell=\frac{\ell_0^2}{\ell},
\label{82}
\end{equation}
in analogy with Eq. (\ref{sesenta}). Last but not least, the continuum limit is recovered from (\ref{81}) as either one of these two cases: $\ell\to 0$ or $\tilde\ell\to 0$. Then one of the two Laplacian terms in (\ref{81}) survives the limit as the standard Laplacian on $\mathbb{R}^D$, while the other one can be dropped and the mass terms add up. The result is the continuum action integral (\ref{508}).

We will substantiate our claims by computing the corresponding Feynman propagator $G^{\rm (QG)}(s)$ and verifying to what extent an equality analogous to (\ref{cientoveintuno}), namely
\begin{equation}
{\cal G}_D^{\rm (QG)}(s)=G_D^{\rm (QG)}(s),
\label{503}
\end{equation}
can hold true. To this end we will use the heat kernel method to derive all propagators. In an appendix (section \ref{sekt5}) we recall the basics of this method as applied to the case when $\ell=0$ (a continuum field theory); for a review see, {\it e.g.}\/, ref. \cite{VASSILEVICH}.

With position space in Eq. (\ref{81}) given by $\mathbb{Z}^D(\ell)\times\mathbb{Z}^D(\tilde\ell)$, let us recall the basics of quantum theory on a lattice; for details we refer to the textbook \cite{SMIT}.

\subsection{Quantum theory on a lattice}\label{sekt7}

The scalar product $\langle f\vert g\rangle$ of functions $f,g$ on $\mathbb{Z}^D(\ell)\times\mathbb{Z}^D(\tilde\ell)$ is defined as a sum over lattice positions:
\begin{equation}
\langle f\vert g\rangle=\sum_{{\bf y}\in\mathbb{Z}^D(\ell)\times\mathbb{Z}^D(\tilde\ell)}f^*({\bf y})g({\bf y})=\sum_{{\bf x}\in\mathbb{Z}^D(\ell)}f^*({\bf x})g({\bf x})
\times\sum_{\tilde {\bf x}\in\mathbb{Z}^D(\tilde\ell)}f^*(\tilde{\bf x})g(\tilde{\bf x}).
\label{26}
\end{equation}
The eigenvalue equation corresponding to the action functional (\ref{81}),
\begin{equation}
\left(-\nabla_{\ell}^2+\frac{m^2}{2}-\nabla_{\tilde {\ell}}^2+\frac{m^2}{2}\right)\Phi_{{\bf p},\tilde {\bf p}}({\bf x}, \tilde{\bf x})=\lambda_{{\bf p},\tilde{\bf p}}(\ell,\tilde\ell)\Phi_{{\bf p},\tilde{\bf p}}({\bf x}, \tilde{\bf x}),
\label{600}
\end{equation}
is solved by
\begin{equation}
\Phi_{{\bf p},\tilde{\bf p}}({\bf x},\tilde{\bf x})=\exp\left[{\rm i}\left({\bf p}\cdot{\bf x}+\tilde {\bf p}\cdot\tilde{\bf x}\right)\right], \quad {\bf x}\in\mathbb{Z}^D(l),\quad \tilde{\bf x}\in\mathbb{Z}^D(\tilde l)
\label{601}
\end{equation}
provided the eigenvalues $\lambda_{{\bf p},\tilde{\bf p}}(\ell,\tilde\ell)$ satisfy the dispersion relation
\begin{equation}
\lambda_{{\bf p},\tilde{\bf p}}(\ell, \tilde\ell)=\frac{m^2}{2}+\frac{2}{\ell^2}\sum_{j=1}^D\left[1-\cos\left(\ell p_j\right)\right]+\frac{m^2}{2}+\frac{2}{\tilde\ell^2}\sum_{j=1}^D\left[1-\cos\left(\tilde \ell \tilde p_j\right)\right],
\label{602}
\end{equation}
where $p_j={\bf p}\cdot{\bf e}_j$ and $\tilde p_j=\tilde{\bf p}\cdot{\bf e}_j$.
As a consistency check, Taylor expanding $\cos z=1-z^2/2+\ldots$ we see that either one of the two limits $\ell\to 0$, $\tilde\ell\to 0$  duly reduces (\ref{602}) to the free--field dispersion relation $\lambda_{\bf p}=p^2+m^2$ valid for the continuum action (\ref{508}).  Powers of momentum higher than quadratic, such as those in Eq. (\ref{602}), are indicative of the effective nature of the theory described by the action functionals (\ref{506}), (\ref{81}). Ultimately this is due to the discretisation of space introduced by the quantum of length, {\it i.e.}\/, by the replacement of the continuous space $\mathbb{R}^D$ with a lattice counterpart.

Due to the cosine in (\ref{602}), all momentum components $p_j$, $\tilde p_j$ are bounded: 
\begin{equation}
p_j\in\mathbb{F}\left(\frac{\pi}{\ell}\right)=\left[-\frac{\pi}{\ell},\frac{\pi}{\ell}\right], \qquad \tilde p_j\in\mathbb{F}\left(\frac{\pi}{\tilde\ell}\right)=\left[-\frac{\pi}{\tilde\ell},\frac{\pi}{\tilde\ell}\right].
\label{30}
\end{equation}
The allowed momenta $({\bf p},\tilde{\bf p})$ are all the vectors within the $2D$--dimensional cube
\begin{equation}
\mathbb{F}^D\left(\frac{\pi}{\ell}\right)\times \mathbb{F}^D\left(\frac{\pi}{\tilde \ell}\right)=\mathbb{F}\left(\frac{\pi}{\ell}\right)\times\ldots\times\mathbb{F}\left(\frac{\pi}{\ell}\right)\times\mathbb{F}\left(\frac{\pi}{\tilde\ell}\right)\times\ldots\times\mathbb{F}\left(\frac{\pi}{\tilde\ell}\right).
\label{607}
\end{equation}
The domain $\mathbb{F}^D\left(\pi/\ell\right)\times\mathbb{F}^D(\pi/\tilde\ell)$ is known as the first Brillouin zone. Its role is to enforce an UV cutoff for momenta.

We can finally write the orthonormal eigenfunctions (\ref{601}) as
\begin{equation}
\Phi_{{\bf p},\tilde{\bf p}}({\bf x},\tilde{\bf x})=\left(\frac{\ell}{2\pi}\right)^{D/2}\left(\frac{\tilde\ell}{2\pi}\right)^{D/2}\exp\left[{\rm i}\left({\bf p}\cdot{\bf x}+\tilde{\bf p}\cdot\tilde{\bf x}\right)\right], 
\label{603}
\end{equation}
Moreover, as we let $\left({\bf p},\tilde{\bf p}\right)\in\mathbb{F}^D(\pi/\ell)\times \mathbb{F}^D(\pi/\tilde\ell)$, the $\Phi_{{\bf p},\tilde{\bf p}}({\bf x},\tilde{\bf x})$ form a complete set of functions on the Hilbert space $L^2\left(\mathbb{Z}^D(\ell)\times\mathbb{Z}^D(\tilde\ell)\right)$ of quantum states.

\subsection{The propagator in momentum space}\label{sekt8}

The position--space representation of the scalar Feynman propagator on the lattice $\mathbb{Z}^D(\ell)$ has been analysed in ref. \cite{IRELAND}; unfortunately there appears to be no tractable analytic expression for this propagator except in certain asymptotic limits. Here we face the same problem on the lattice $\mathbb{Z}^D(\ell)\times\mathbb{Z}^D(\tilde\ell)$. We will therefore tackle the computation in momentum space, where things simplify somewhat, and determine an asymptotic expression for 
$\hat G_D^{\rm (QG)}({\bf q},\tilde{\bf q},{\bf p},\tilde{\bf p})$. Our objective is to determine to what extent Eq. (\ref{503}) can hold, whereby the left--hand side ${\cal G}_D^{\rm (QG)}(s)$ is known by Eq. (\ref{kagonesteve}) but the right--hand side $G_D^{\rm (QG)}(s)$ remains unknown. 

To begin with, for the heat kernel ${\cal K}({\bf q},\tilde{\bf q}; {\bf p},\tilde{\bf p};\tau)$ corresponding to the Helmholtz operator in the action (\ref{81}) one obtains
$$
{\cal K}({\bf q},\tilde{\bf q}; {\bf p},\tilde{\bf p};\tau)=\langle{\bf q},\tilde{\bf q}\vert\exp\left(-\nabla_{\ell}^2+\frac{m^2}{2}-\nabla_{\tilde {\ell}}^2+\frac{m^2}{2}\right)
\vert{\bf p},\tilde{\bf p}\rangle
$$
\begin{equation}
=\exp\left[-\tau\lambda_{{\bf p},\tilde{\bf p}}(\ell,\tilde\ell)\right]\delta\left({\bf q}-{\bf p}\right)\delta\left(\tilde{\bf q}-\tilde{\bf p}\right).
\label{605}
\end{equation}
This is formally identical to its counterpart on $\mathbb{R}^D$ (Eq. (\ref{22}) of the appendix), albeit with a different set of eigenvalues. Next we integrate (\ref{605}) with respect to the spectral parameter $\tau$ in order to obtain the propagator $\hat G_D^{\rm (QG)}({\bf q},\tilde{\bf q},{\bf p},\tilde{\bf p})$ in momentum space:
\begin{equation}
\hat G_D^{\rm (QG)}({\bf q},\tilde{\bf q},{\bf p},\tilde{\bf p})=\int_0^{\infty}{\rm d}\tau\,{\cal K}({\bf q},\tilde{\bf q},{\bf p},\tilde{\bf p};\tau)=\frac{1}{\lambda_{{\bf p},\tilde{\bf p}}(\ell,\tilde\ell)}\delta\left({\bf q}-{\bf p}\right)\delta\left(\tilde{\bf q}-\tilde{\bf p}\right).
\label{610}
\end{equation}
This is again identical to its counterpart on $\mathbb{R}^D$ (Eq. (\ref{23}) of the appendix), only the eigenvalues are different. 

On the other hand, the Fourier transform of the point--particle propagator (\ref{kagonesteve}) reads
\begin{equation}
\hat{\cal G}_D^{\rm (QG)}(p)=\frac{m^{D-2}}{(2\pi)^{D/2}}\,{\cal F}\left[\frac{K_{D/2-1}\left(m\sqrt{s^2+\ell^2}\right)}{\left(m\sqrt{s^2+\ell^2}\right)^{D/2-1}}\right]=\ell\frac{K_1\left(\ell\sqrt{p^2+m^2}\right)}{\sqrt{p^2+m^2}}.
\label{90}
\end{equation}
To gain some insight we expand $K_1(z)\simeq 1/z+O(z^0)$. As $\ell\to 0$, Eq. (\ref{90}) becomes
\begin{equation}
\hat{\cal G}_D^{\rm (QG)}(p)\simeq\frac{1}{p^2+m^2}+O(\ell).
\label{92}
\end{equation}
This is perfectly fine, as it equals the momentum--space Feynman propagator for a free particle. We need to check (\ref{92}) against its lattice counterpart (\ref{610}), with the eigenvalue (\ref{602}). When $\ell\to 0$ we can Taylor--expand (\ref{602}) in $\ell$ to obtain
$$
\lambda_{{\bf p},\tilde{\bf p}}(\ell,\tilde\ell)\simeq p^2+m^2-\frac{\ell^2}{12}\sum_{j=1}^Dp_j^4+ \frac{2}{\tilde\ell^2}\sum_{j=1}^D\left[1-\cos\left(\tilde \ell \tilde p_j\right)\right]
$$
\begin{equation}
\simeq p^2+m^2-\frac{\ell^2}{12}\sum_{j=1}^Dp_j^4,
\label{60}
\end{equation}
because $\tilde\ell\to\infty$ as $\ell\to 0$. Further dropping $O(\ell^2)$ terms establishes the equality between Eqs. (\ref{610}) and (\ref{92}) to $O(\ell^0)$. That is, the momentum--space, lattice field--theory propagator $\hat G_D^{\rm (QG)}$ and its point--particle counterpart $\hat{\cal G}_D^{\rm (QG)}$ coincide when $\ell=0$, as had to be the case for consistency.

However, $O(\ell^2)$ terms derived from Eq. (\ref{90}) do {\it not}\/ match the $\sum_{j=1}^Dp_j^4$ term in Eq. (\ref{60}). As announced, our lattice model must be discarded as a field--theory equivalent of a point particle in the presence of a quantum of length.

\section{Discussion}\label{sekt10}

For as long as an established theory of quantum gravity (QG) is lacking, a reasonable course to take is to study the mesoscopics of such an eventual theory. That is, one acknowledges one's ignorance about the ultimate structure of spacetime, and instead tries to concentrate on first--order, quantum--gravitational effects that our current theories might still be able to probe. It is generally accepted that one such effect is the existence of a quantum of length $\ell$. Indeed the latter plays a key role in numerous approaches to quantum gravity (see, {\it e.g.}\/, refs. \cite{GARAY, HOSSENFELDER, KIEFER1}). The quantum of length $\ell$ can be identified with (possibly some multiple of) the Planck length $L_P$. The mesoscopic regime would then correspond to lengths that would be small multiples of the quantum of length.

An example will help to clarify the notion of a {\it mesoscopic effect}\/. The kinetic theory of matter, while operating on the basis that the latter is granular, made no assumptions regarding the specific nature of the atoms. Still, the realisation that Avogadro's number plays a key role is a success of the kinetic theory. The mere existence of Avogadro's number can be regarded as a mesoscopic effect of the atomic theory of matter, an effect that was tested using the less accurate kinetic theory.

Quantum--gravity effects due to a quantum of length have been analysed in the literature under a wide array of different viewpoints; a very incomplete sample might include the following. Historically, the first formal attempt to quantise spacetime itself dates back to Snyder's groundbreaking paper \cite{SNYDER}. More recently, refs. \cite{ROY1, ROY2, FINSTER, SINGH1} deal with different manifestations of the zero--point length in quantum gravity and cosmology. The existence of an UV cutoff in momentum space has implications on the geometrical properties of spacetime and the Feynman propagators defined on them, as evidenced in ref. \cite{KOTHAWALA1}. The generalised uncertainty principle plays a key role in approaches to quantum field theory in the presence of a quantum of length \cite{BOSSO, CASADIO, MONDAL1, MONDAL2}. The microscopic structure of spacetime, as well as some of its properties from the point of view of themodynamics and statistical mechanics, have been the subject of refs. \cite{PADDY5, PESCI1,PESCI2, PESCI3}. 

In this article we have taken Padmanabhan's works \cite{PADDY0, PADDY1} on the quantum of length as our starting point. In the absence of a quantum of length there is an equivalence between the point--particle description (\ref{linint}) and the field--theory description (\ref{508}). In the presence of a quantum of length there is a point--particle description given by Eq. (\ref{treintaydos}); we have asked ourselves what the field--theory analogue of the point--particle theory (\ref{treintaydos}) might be. Restricting ourselves to terms up to $O(\ell^2)$, this field--theory analogue turns out to be a scalar field theory defined in $D+2$ dimensions,  where $D$ is the dimensionality of the space where the point particle lives. 

Specifically, the field--theory action functional we put forward is (\ref{93}); the dimensional extension from 
$\mathbb{R}^D$ to $\mathbb{R}^{D+2}$ is in perfect parallel with the point--particle approach followed in ref. \cite{PADDY6}. Compared to the analysis in ref. \cite{JAPON2}, our action functional (\ref{93}) contains just two derivatives in position space, as opposed to the infinite tower of space derivatives in \cite{JAPON2}. The action integral in \cite{JAPON2} is defined on a continuum spacetime, but the heat kernel gets discretised. 

Eq. (\ref{93}) is the key point of our construction. It implies that, up to $O(\ell^2)$, QG corrections to a field theory defined in 
$\mathbb{R}^D$ can be modelled via a QG--free field--theory defined in $\mathbb{R}^{D+2}$. Typically a quantum of length $\ell$ introduces nonlocal operators. Our Eq. (\ref{93}) bypasses this nonlocality by dimensional extension, as in ref. \cite{PADDY6}. The $(D+2)$--dimensional theory remains local since it contains just two derivatives. The two extra dimensions in $\mathbb{R}^{D+2}$ provide the necessary room for QG fluctuations to play out.

Some steps in the present paper are heuristic and meant as an effective approach to Padmanabhan’s duality--invariant propagator. For example, one would like to have a derivation from a fundamental QG path integral. The numerical coefficient in Eq. (\ref{93}) is fixed by a pragmatic matching rather than as the result of an explicit integration of microscopic degrees of freedom.  The size of higher--order corrections ({\it i.e.}\/, beyond $O(\ell^2)$) remains to be estimated. Resolving the items above is left for future work.

\section{Appendix: propagators from the heat kernel}\label{sekt5}

Given the continuum field theory (\ref{508}), the heat kernel ${\cal K}({\bf x},{\bf y};\tau)$ of the Helmholtz operator $-\nabla_{\bf x}^2+m^2$ satisfies the heat equation
\begin{equation}
\left(-\nabla^2_{\bf x}+m^2\right){\cal K}({\bf x},{\bf y};\tau)+\frac{\partial}{\partial\tau}{\cal K}({\bf x},{\bf y};\tau)=0,\qquad \tau\in(0,\infty)
\label{5}
\end{equation}
subject to the initial condition ${\cal K}({\bf x},{\bf y};0)=\delta({\bf x}-{\bf y})$. One integrates ${\cal K}({\bf x},{\bf y};\tau)$ over the spectral parameter $\tau\in(0,\infty)$ in order to arrive at the Feynman propagator:
\begin{equation}
G_D({\bf x},{\bf y})=\int_0^{\infty}{\rm d}\tau\,{\cal K}({\bf x},{\bf y};\tau).
\label{4}
\end{equation}
Indeed, assuming 
\begin{equation}
\lim_{\tau\to\infty}{\cal K}({\bf x},{\bf y};\tau)=0,
\label{8}
\end{equation}
we have
$$
\left(-\nabla^2_{\bf x}+m^2\right)G_D({\bf x},{\bf y})=\int_0^{\infty}{\rm d}\tau\,\left(-\nabla^2_{\bf x}+m^2\right){\cal K}({\bf x},{\bf y};\tau)
$$
\begin{equation}
=-\int_0^{\infty}{\rm d}\tau\,\frac{\partial}{\partial\tau}{\cal K}({\bf x},{\bf y};\tau)={\cal K}({\bf x},{\bf y}; 0)=\delta\left({\bf x}-{\bf y}\right)
\label{7}
\end{equation}
as desired. 

To determine ${\cal K}({\bf x},{\bf y};\tau)$ one first computes the matrix elements 
\begin{equation}
{\cal K}({\bf x},{\bf y};\tau)=\langle{\bf x}\vert \exp\left[-\tau\left(-\nabla^2_{\bf y}+m^2\right)\right]\vert{\bf y}\rangle
\label{1}
\end{equation}
between position eigenstates. Next one inserts a resolution of the identity in terms of momentum eigenstates $\vert{\bf p}\rangle$. For concreteness assume this resolution to be\footnote{The integral (\ref{9}) might stand for a sum, and it might also by multiplied by some normalisation factor, omitted here for generality. By the same token, the integration range over the momenta is left unspecified.}
\begin{equation}
\int{\rm d}^D{\bf p}\,\vert{\bf p}\rangle\langle{\bf p}\vert={\bf 1}.
\label{9}
\end{equation}
Then
\begin{equation}
{\cal K}({\bf x},{\bf y};\tau)=\int{\rm d}^D{\bf p}\,\langle{\bf x}\vert{\bf p}\rangle \exp\left[-\tau\left(-\nabla^2_{\bf y}+m^2\right)\right]\langle{\bf p}\vert{\bf y}\rangle.
\label{10}
\end{equation}
It will be convenient to take the wavefunctions $\langle{\bf y}\vert{\bf p}\rangle=\Phi_{\bf p}({\bf y})$ as eigenfunctions of the operator $-\nabla^2_{\bf y}+m^2$:
\begin{equation}
\left(-\nabla^2_{\bf y}+m^2\right)\Phi_{\bf p}({\bf y})=\lambda_{\bf p}\Phi_{\bf p}({\bf y}).
\label{2}
\end{equation}
The corresponding eigenvalues $\lambda_{\bf p}$ will satisfy a certain dispersion relation to be determined presently. In this way the heat kernel (\ref{10}) becomes
\begin{equation}
{\cal K}({\bf x},{\bf y};\tau)=\int{\rm d}^D{\bf p}\,\Phi_{\bf p}({\bf x})\Phi^*_{\bf p}({\bf y})\exp\left(-\tau\lambda_{\bf p}\right),
\label{12}
\end{equation}
and the Feynman propagator (\ref{4}) reads
\begin{equation}
G_D({\bf x},{\bf y})=\int_0^{\infty}{\rm d}\tau\,\int{\rm d}^D{\bf p}\,\Phi_{\bf p}({\bf x})\Phi^*_{\bf p}({\bf y})\exp\left(-\tau\lambda_{\bf p}\right).
\label{13}
\end{equation}
The above results are often more conveniently expressed in momentum space. For reference we recall
\begin{equation}
{\cal K}({\bf p},{\bf q};\tau)=\langle{\bf p}\vert\exp\left[-\tau\left(-\nabla^2+m^2\right)\right]\vert{\bf q}\rangle=\exp\left(-\tau\lambda_{\bf p}\right)\delta\left({\bf p}-{\bf q}\right)
\label{22}
\end{equation}
and  
\begin{equation}
G_D({\bf p},{\bf q})=\int_0^{\infty}{\rm d}\tau\,{\cal K}({\bf p},{\bf q};\tau)=\frac{1}{\lambda_{\bf p}}\delta\left({\bf p}-{\bf q}\right).
\label{23}
\end{equation}

In the particular case of Euclidean $\mathbb{R}^D$, the solution to the eigenvalue problem (\ref{2}) is given by the (orthonormal, complete set of) eigenfunctions
\begin{equation}
\Phi_{\bf p}({\bf x})=\frac{1}{(2\pi)^{D/2}}\exp\left({\rm i}{\bf p}\cdot{\bf x}\right), \qquad \lambda_{\bf p}=p^2+m^2,\qquad {\bf p}\in\mathbb{R}^D,
\label{3}
\end{equation}
and substitution of (\ref{3}) into (\ref{12}) produces the heat kernel
$$
{\cal K}({\bf x},{\bf y};\tau)=\frac{1}{(2\pi)^D}\int_{\mathbb{R}^D}{\rm d}^D{\bf p}\exp\left[-{\rm i}{\bf p}\cdot({\bf y}-{\bf x})-\tau\left(p^2+m^2\right)\right]
$$
\begin{equation}
=\frac{1}{(4\pi\tau)^{D/2}}\exp\left(-m^2\tau-\frac{s^2}{4\tau}\right).
\label{536}
\end{equation}
For the Feynman propagator (\ref{13}) we finally find
$$
G_D({\bf x},{\bf y})=\frac{1}{(2\pi)^D}\int_0^{\infty}{\rm d}\tau\,\int_{\mathbb{R}^D}{\rm d}^D{\bf p}\,\exp\left[-\tau\left(p^2+m^2\right)-{\rm i}{\bf p}\cdot\left({\bf y}-{\bf x}\right)\right]
$$
\begin{equation}
=\int_{\mathbb{R}^D}\frac{{\rm d}^D{\bf p}}{(2\pi)^D}\frac{\exp\left[{\rm i}{\bf p}\cdot\left({\bf x}-{\bf y}\right)\right]}{p^2+m^2},
\label{14}
\end{equation}
in nice agreement with our previous result (\ref{trescientoscuatro}).

\vskip1cm
\noindent
{\bf Acknowledgments}\\ 
The authors acknowledge support by project MCIN/AEI/10.13039/501100011033 and by ERDF, “A way of making Europe”, under project PID2024-162480OB-I00.

\end{document}